\documentclass{aastex631}
\usepackage{bm}
\usepackage{threeparttable}
\usepackage{graphicx}
\usepackage{hyperref}
\hypersetup{
  colorlinks=true,
  urlcolor=blue,
}

\makeatletter

\newcommand{\Rmrum}[1]{\expandafter\@slowromancap\romannumeral #1@}
\makeatother

\usepackage{amsmath}
\usepackage{amsthm}
\usepackage{amssymb}
\usepackage{mathtools, nccmath}
\usepackage{wrapfig}
\usepackage{comment}
\usepackage{blindtext}
\usepackage{xspace}
\usepackage{verbatim}

\shortauthors{Lai et al.}

\submitjournal{PSJ}

\begin{document}

\title{Ocean Circulation on Tide-locked Lava Worlds, Part II: Scalings}

\author[0000-0001-9700-9121]{Yanhong Lai}
\affiliation{Laboratory for Climate and Ocean-Atmosphere Studies, Department of Atmospheric and Oceanic Sciences, School of Physics, Peking University, Beijing 100871, China}
\affiliation{Institute of Ocean Research, Peking University, Beijing 100871, China}

\author[0000-0002-4615-3702]{Wanying Kang}
\affiliation{Department of Earth, Atmosphere and Planetary Science, MIT, Cambridge, MA 02139, USA}
\correspondingauthor{Wanying Kang}
\email{wanying@mit.edu}

\author[0000-0001-6031-2485]{Jun Yang}
\affiliation{Laboratory for Climate and Ocean-Atmosphere Studies, Department of Atmospheric and Oceanic Sciences, School of Physics, Peking University, Beijing 100871, China}
\affiliation{Institute of Ocean Research, Peking University, Beijing 100871, China}

\begin{abstract}

On tidally locked lava planets, magma ocean can form on the permanent dayside. The circulation of the magma ocean can be driven by stellar radiation and atmospheric winds. The strength of ocean circulation and the depth of the magma ocean depend on external forcings and the dominant balance of the momentum equation. In this study, we develop scaling laws for the magma ocean depth, oceanic current speed, and ocean heat transport convergence driven by stellar and wind forcings in three different dynamic regimes: non-rotating viscosity-dominant Regime I, non-rotating inviscid limit Regime II, and rotation-dominant Regime III. Scaling laws suggest that magma ocean depth, current speed, and ocean heat transport convergence are controlled by various parameters, including vertical diffusivity/viscosity, substellar temperature, planetary rotation rate, and wind stress. In general, scaling laws predict that magma ocean depth ranges from a few meters to a few hundred meters. For Regime I, results from scaling laws are further confirmed by numerical simulations. Considering the parameters of a typical lava super-Earth, we found that the magma ocean is most likely in the rotation-dominant Regime III.

\end{abstract}


\section{Introduction}
\label{sec:intro} 

On tidally locked lava planets, magma ocean can form on the permanent dayside \citep{leger2009transiting,leger2011extreme,batalha2011kepler,demory2011detection,dumusque2014kepler,kite2016atmosphere,bourrier201855,malavolta2018ultra,nguyen2020modelling,chao2021lava,brinkman2023toi}. 
Given the high surface temperatures and the relatively advanced ages of lava worlds \citep{leger2009transiting,batalha2011kepler}, there is a possibility that volatile elements, such as C, N, H, may have dissipated from these planets \citep{valencia2010composition}. The atmospheric pressure, determined by the vaporization of underlying silicate melts \citep{schaefer2009chemistry,leger2009transiting,kite2016atmosphere}, may exhibit significant gradients due to the diminishing stellar flux away from the substellar point \citep{leger2011extreme,nguyen2020modelling}. Thus, there are strong horizontal winds flowing outward from the substellar point ($\sim$2000 m\,s$^{-1}$; \cite{castan2011atmospheres,nguyen2020modelling, kang2021escaping}).

Much remains unclear regarding the ocean on tidally locked lava planets. In most of previous studies, the magma ocean depth is assumed to be determined by the adiabatic temperature profile \citep{leger2011extreme,boukare2022deep}. This assumption holds particularly true in scenarios within which a potent internal heat source dominates over stellar radiation \citep{boukare2022deep}. The prevalence of strong internal heating can sustain vigorous vertical convection, leading to an isentropic vertical temperature profile for which temperature increases with pressure following the adiabatic lapse rate \citep{solomatov2007magma,zhang2022internal}. Consequently, the magma ocean depth can extend to tens or even hundreds of kilometers \citep{leger2011extreme,boukare2022deep}. 
 
Even in the presence of internal heating, vigorous convection may not occur within the magma ocean, unless 1) the resultant temperature increase with depth is greater than the increase of liquidus with pressure, or 2) the resultant temperature increase leads to a supercritical Rayleigh number (a dimensionless number that describes the relationship between buoyancy and dissipative forces; \cite{solomatov2007magma}) even in the solid phase.  In the former situation, a second liquid layer would form in the interior, subject to convection, and in the latter situation, the solid mantle will convect. \cite{meier2023} utilized 2D mantle convection models to investigate the magma ocean depth of 55 Cnc e under varying internal heating rates. Their findings suggest that the depth of the magma ocean could still reach approximately 500 km on the dayside, even without internal heating. However, it is important to note that their study did not account for the impact of ocean circulation and the resultant stratification on magma ocean depth.

The magma ocean depth will be controlled by ocean circulation when the internal heat source is weak or absent. This circumstance is plausible for tidally locked lava planets, particularly those that have undergone cooling over billions of years (such as CoRot-7b, Kepler-10b, 55 Cnc e; \cite{leger2011extreme,batalha2011kepler,von201155,malavolta2018ultra,brinkman2023toi}).
Ocean circulation is conceivable on lava planets, considering that the viscosity of fully molten silicates is comparable to that of seawater on Earth \citep{dingwell2004viscosity,haynes2014crc,sun2020physical,zhang2022internal}.
With a substantial surface temperature gradient and robust atmospheric winds \citep{leger2011extreme,castan2011atmospheres, kang2021escaping}, both thermal and wind forcings can play crucial roles in driving the ocean circulation on lava planets.

The planetary parameters of lava planets and the oceanic parameters of magma oceans could vary significantly, implying that magma oceans on these lava planets can exist in different dynamic regimes. Tidally locked lava planets, such as Kepler-10b, CoRot-7b, 55 Cnc e, and K2-141b, typically have relatively fast rotation periods, around one Earth day \citep{leger2011extreme,batalha2011kepler,bourrier201855,malavolta2018ultra,brinkman2023toi}. Magma ocean may also form on slowly rotating terrestrial planets due to giant impacts \citep{elkins2012magma,chao2021lava}.

This paper is Part II of a series of papers. In Part I, we investigated the ocean circulation on tidally locked lava planets using an idealized two-dimensional (2D) numerical model, developed by the authors (Ocean Circulation on Tide-locked Lava Worlds, Part I: An Idealized 2D Numerical Model). We presented the simulation results under thermal forcing only and under both thermal and wind forcings. We further compared our simulation results with previous studies. In general, simulation results under weak or no internal heat source suggest an ocean depth on the order of 100 m, which is over 100 times shallower than that without ocean circulation. However, Coriolis force is neglected in Part I, which could play a significant role in ocean circulation \citep{kite2016atmosphere,vallis2019essentials}.

\begin{figure*}
    \centering
    \includegraphics[width=0.8\textwidth]{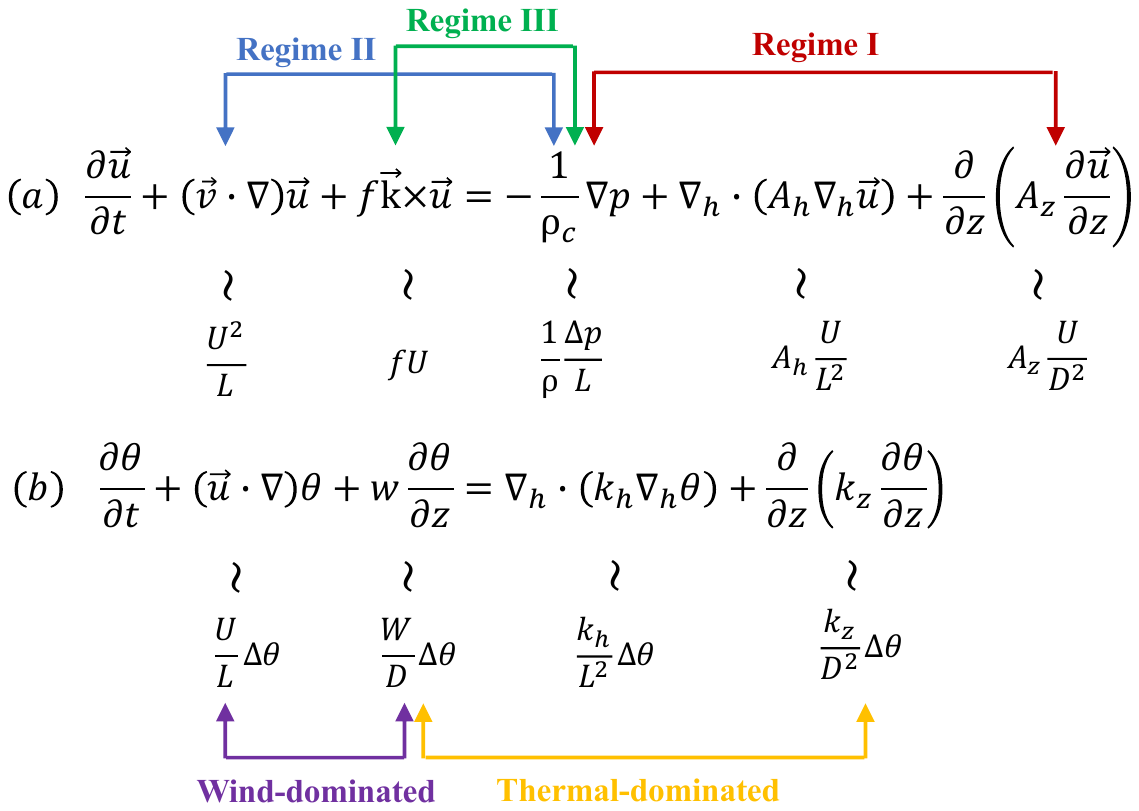}
    \caption{The general equations for horizontal momentum (a) and potential temperature (b) in the interior ocean, where $\textbf{v}=u\textbf{i}+v\textbf{j}+w\textbf{k}$, u, v, and w are zonal, meridional, and vertical velocities, respectively, $\textbf{u}=u\textbf{i}+v\textbf{j}$ is horizontal velocity, $f$ is Coriolis parameter, $\rho_c$ is reference density, $p$ is pressure, $A_h$ and $A_z$ are horizontal and vertical viscosity coefficients, respectively, 
    $\theta$ is potential temperature, $k_h$ and $k_z$ are horizontal and vertical diffusivities, respectively. The characteristic scales of each term are presented below the equations. The dominant terms of momentum equation in Regimes I, II, and III are shown in red, blue, and green brackets, respectively. 
    The dominant terms of thermal equation in thermal-dominated and wind-dominated systems are shown in gold and purple brackets, respectively.}
    \label{fig:equation}
\end{figure*}

Here, we present scaling laws for magma ocean depth, oceanic current speed, and ocean heat transport convergence in three dynamical regimes (Figure \ref{fig:equation}): non-rotating viscosity-dominant Regime I in Section \ref{scaling_2D}, non-rotating inviscid limit Regime II in Section \ref{scaling_invis}, and rotation-dominant Regime III in Section \ref{scaling_3D}. We present the comparison between the results from scaling laws and numerical simulations in Regime I in Section \ref{scaling_2D}. We discuss the parameter space within which each of the scaling laws is applicable in Section \ref{sec:parameter_regime}. In Section \ref{sec:conclude}, conclusions are presented.

\section{Scaling laws for magma ocean depth, current speed, and ocean heat transport convergence}\label{sec:theoretical}


There are three major driving forces in the system: 1) thermal forcing from the star, 2) wind stress induced by the day- to night-side atmospheric flow, and 3) the surface evaporation and condensation. Our work here focuses on the first two types of forcing. The third one is left for future study.

On lava worlds, the stellar radiative flux is in the order of $10^6$~W\,m$^{-2}$, which has been shown to be 2-3 orders of magnitude greater than the heat rates induced by other physical processes, such as advection \citep{kite2016atmosphere,kang2021escaping}. The dominance of stellar radiative flux justifies our setting the surface temperature to the radiative equilibrium temperature. 
The wind stress, i.e., the momentum exchange between the atmosphere and ocean, is induced by the deposition of atmospheric mass with non-zero velocity, and momentum transport by turbulent eddies \citep{ingersoll1985supersonic}. On tidally locked lava planets, eddy transport dominate given that the wind speed $V_a$ can reach $\sim$2000 m\,s$^{-1}$ \citep{castan2011atmospheres,nguyen2020modelling,kang2021escaping}. Only accounting for this component, the wind stress on such planets can be estimated as $\tau = \rho_a C_D V_a^2$, where $\rho_a \sim 10^{-2}$ kg\,m$^{-3}$ represents atmospheric density \citep{kang2021escaping}, and $C_D$ represents the surface drag coefficient. Assuming $C_D \sim 10^{-2}$ under high wind speeds \citep{sterl2017drag,jiang2021high}, the wind stress is approximately 100 N\,m$^{-2}$. More details of wind stress calculation can be found in Part I of this series of paper.

\begin{figure*}
    \centering
    \includegraphics[width=0.85\textwidth]{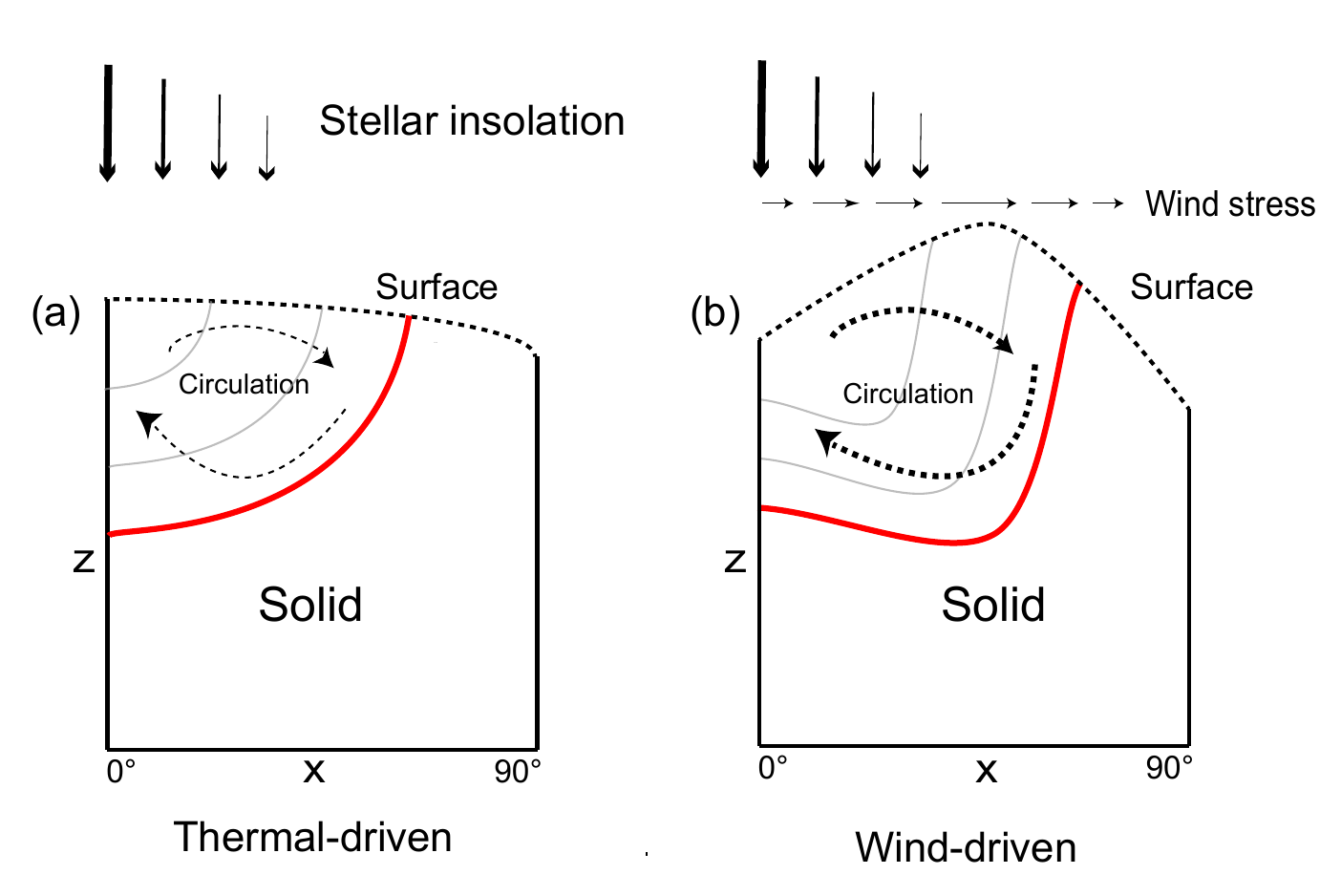}
    \caption{Schematics of thermal-driven (a) and wind-driven (b) ocean circulations. In the schematics, the external forcings, surface height, ocean circulation, and magma ocean depth are shown. The red solid line represents the boundary of the magma ocean, and the grey lines represent the isotherms, indicating that temperature decreases as depth increases within the ocean. For more details, please see Part I.}
    \label{fig:schematic}
\end{figure*}

When the stellar radiative forcing dominates (see Figure \ref{fig:schematic}(a)), the magma ocean circulation is mainly powered by the density gradient induced by the attenuation of stellar flux away from the substellar point. Under this forcing, the fluid density increases with $x$, the distance away from the substellar point, driving an ocean current toward the substellar point in the lower part of the ocean. In an equilibrium state, net transport across different depth $z$ needs to vanish, so the surface fluid needs to flow toward the edge of the magma ocean, facilitated by the pressure gradient induced by the sea surface elevation anomaly. 

When the wind stress dominates (see Figure \ref{fig:schematic}(b)), the surface fluid will be pushed toward the magma ocean edge, creating a high sea surface height there. This high sea surface height then forces fluid back toward the substellar point in the lower levels. Although the circulation in the thermal- and wind-driven systems both features sinking motion along the magma ocean edge and upwelling motions near the substellar point, the shape of the sea surface and the liquid-rock boundary differs. 

In the presence of ocean circulation, the ocean temperature profile differs from that without such circulation \citep{leger2011extreme,boukare2022deep}. 
Since both forcings cause cold magma formed at the magma ocean edge to sink, filling the bottom ocean, temperature in the ocean should decrease with depth from the surface temperature to the liquidus (Figure \ref{fig:schematic}). Below the ocean bottom, vertical diffusion governs the temperature profile, which should be uniform at a temperature just below the liquidus. For more details, please see Part I of our work. Here, we focus on the magma ocean depth determined by the ocean circulation.

As to be shown later, vertical diffusivity and viscosity play a key role in determining the magma ocean circulation strength and depth, by controlling the rate at which dense fluid can be pumped back up to the surface and the rate momentum can be transported vertically. 
Diffusivity and viscosity of molten silicates are contributed by both molecular random motions and turbulent fluid motions. While molecular diffusivity varies from 10$^{-9}$ to 10$^{-8}$ m$^2$\,s$^{-1}$ at a temperature of 2000-3000 K \citep{ghosh2011diffusion} and molecular viscosity is around 10$^{-4}$ m$^2$\,s$^{-1}$ \citep{sun2020physical,zhang2022internal}, turbulence-induced diffusivity/viscosity powered by winds and tides remains largely unconstrained. In Earth's oceans, vertical mixing is mainly contributed by turbulence, yielding a vertical diffusivity around $10^{-5}$--$10^{-3}$~m$^2$\,s$^{-1}$ \citep{munk1998abyssal,Waterhouse2014} and a vertical viscosity around $10^{-3}$--$10^{-1}$ m$^2$\,s$^{-1}$ \citep{luo2023study,sentchev2023estimation}, several orders of magnitude greater than the molecular diffusivity/viscosity. Given the large uncertainty, we explore a wide range of diffusivity and viscosity values, ranging from 10$^{-7}$ to 10$^{-3}$ m$^2$\,s$^{-1}$ and 10$^{-7}$ to 10$^{2}$ m$^2$\,s$^{-1}$, respectively.  

In this section, we present scaling laws for the horizontal velocity, magma ocean depth, and ocean heat transport convergence in three dynamical regimes that feature different dominant balances in momentum equation: non-rotating viscosity-dominant Regime I (Section \ref{scaling_2D}), non-rotating inviscid limit Regime II (Section \ref{scaling_invis}), and rotation-dominant Regime III (Section \ref{scaling_3D}). In Regimes I, II, and III, the pressure gradient force is balanced by the vertical viscosity, nonlinear advection, and Coriolis force in the momentum equation, respectively. Figure \ref{fig:equation}(a) shows the general horizontal momentum equation and marks the dominant terms in each dynamical regime.
Note that geometry introduces additional metric terms into the equations. However, from a scaling perspective, these metric terms do not affect the final results up to a constant factor.
Also, it has been shown that the influence of varying coordinates on the dynamics of tidally locked planets might be limited \citep{hammond2018wave,wang2021phase,yang2023cloud}. 

For each dynamic regime, the predominant terms in the thermal equation vary depending on whether ocean circulation is predominantly driven by thermal or wind forcings \citep{vallis2019essentials}.
When wind forcing is absent or weak, vertical advection and vertical diffusion take precedence over other terms in the thermal equation (gold bracket in Figure \ref{fig:equation}(b)). Conversely, when wind forcing is significant, horizontal advection and vertical advection become dominant (purple bracket in Figure \ref{fig:equation}(b)).
Thus, we provide two scaling laws in each dynamic regime, which describe the magma ocean depth $D$, horizontal velocity $U$, and ocean heat transport convergence $H$ dominated by thermal and wind forcings, respectively.
Knowing the strengths of the two forcings, one could evaluate $D$, $U$, and $H$ following the two scaling laws, and choose whichever that gives the larger value for $D$, $U$ and $H$. This will guarantee the scaling law for the dominant type of forcing get selected.


\subsection{Regime I: non-rotating viscosity-dominant regime}\label{scaling_2D}

In the non-rotating regime, Coriolis force is neglected, which guarantees the azimuthal symmetry of the circulation, i.e., the dynamics only varies with depth $z$ and distance from the substellar point $x$. If further the vertical viscosity coefficient is large, the momentum budget will be in balance between the pressure gradient force and vertical viscosity (red brackets in Figure \ref{fig:equation}(a)).

The dominant balance of momentum equation gives
\begin{equation}\label{eq:momentum_balance1}
    \frac{1}{\rho_c} \frac{\partial p}{\partial x} \approx \frac{\partial}{\partial z}\left(A_z \frac{\partial u}{\partial z}\right),
\end{equation}
where $u$ is horizontal velocity, $\rho_c$ is reference density, $A_z$ is vertical viscosity coefficient, and $p = \rho_c g \eta - \rho_c \int_z^0 b dz$ is pressure, where $b = -g\frac{\rho}{\rho_c}$ is buoyancy, $g$ is gravity, $\rho$ is density, and $\eta$ is sea surface height (SSH). Near the surface, pressure gradient is dominated by the SSH variation, driving a flow away from the substellar point, whereas the pressure gradient near the bottom of the ocean should be dominated by the integrated buoyancy anomaly, driving a return flow toward the substellar point. For pressure gradient to reverse in the vertical direction, the integrated buoyancy anomaly should be of the same magnitude as the pressure gradient.
It should be noted that, in our study, the magma ocean boundary is set by the liquidus, where the magma viscosity is generally low, so that vertical viscosity $A_z$ may be considered as a constant throughout our domain. 

In the interior ocean, mass continuity is always satisfied,
\begin{equation}
\label{eq:mass_continuity}
    \frac{\partial u }{\partial x}+\frac{\partial w}{\partial z}=0,
\end{equation}
where $w$ denotes vertical velocity.
When wind forcing is weak, the dominant driver of the circulation is diffusion. Along the magma ocean edge, fluid cools and sinks to the bottom, releasing gravity potential energy. 
In a stably stratified ocean, vertical mixing induced by tidal and wind-driven wave breaking can pump denser fluid upward, thereby increasing the gravitational potential energy of the system \citep{wunsch2004}.  
In the upwelling regions near the substellar point, the downward diffusive heat/buoyancy flux needs to balance the upward advective heat/buoyancy flux, leading to the so called advective-diffusive balance (the gold bracket in Figure \ref{fig:equation}(b))
\begin{equation}
\label{eq:advective_diffusive_balance}
    w\frac{\partial \theta}{\partial z}=\kappa_z\frac{\partial^2 \theta}{\partial z^2},
\end{equation}
where $\theta$ denotes potential temperature, and $k_z$ denotes vertical diffusivity.

Thus, the scales of Equation (\ref{eq:momentum_balance1}), (\ref{eq:mass_continuity}), and (\ref{eq:advective_diffusive_balance}) could be written as,
\begin{equation}\label{eq:scaling_momentum_balance1}
    \frac{\Delta b }{L} D \sim A_z \frac{U}{D^2}, 
\end{equation}
\begin{equation}\label{eq:scaling_mass_continuity}
    \frac{U}{L} \sim \frac{W}{D}, 
\end{equation}
\begin{equation}\label{eq:scaling_advection_diffusion}
     D \sim \frac{k_z}{W},
\end{equation}
where $U$, $W$, $L$, $D$, and $\Delta b$ represent the typical scales of horizontal velocity, vertical velocity, horizontal range of the magma ocean, magma ocean depth, and horizontal buoyancy difference, respectively. Especially, $\Delta b \sim g\alpha \Delta T$ utilizing the linear equation of state (EoS) approximation, where $\alpha$ is thermal expansion coefficient, and $\Delta T$ is the horizontal potential temperature difference within the ocean. 
Note that the mass continuity equation (Equation (\ref{eq:mass_continuity})) is written in a Cartesian coordinate. If it is rewritten for a disk coordinate, the first term of Equation (\ref{eq:mass_continuity}) will change to $\frac{1}{x}(\frac{\partial ux}{\partial x})$, but this change will not affect the scaling presented in Equation (\ref{eq:scaling_mass_continuity}).

\begin{table}
\caption{Baseline Values of Planetary and Oceanic Parameters \label{tab1}}
\centering
\begin{tabular}{lll}
\hline
Parameter  & Value  & Units\\
\hline
Planet radius ($a$) & 10$^7$ & m\\
Planet gravity ($g$) & 20 & m\,s$^{-2}$\\
Planet rotation rate ($\Omega$) & 0 & s$^{-1}$\\
Substellar temperature ($T_{sub}$) & 3000 & K\\
Liquidus ($T_{liq}$) & 2000 & K\\
Temperature contrast within the ocean ($\Delta T = T_{sub} - T_{liq}$) & 1000 & K\\
Angular radius of the ocean ($\theta_p \approx cos^{-1}\left(\frac{T_{liq}}{T_{sub}}\right)$)  & 2 & \\
Horizontal scale of the ocean ($L \approx a \theta_p$)  & 2$\times$10$^7$ & m\\
Thermal expansion coefficient ($\alpha$) & 2$\times 10^{-4}$ & K$^{-1}$\\
Heat capacity at constant pressure ($c_p$) & 1800 & J\,kg$^{-1}$\,K$^{-1}$\\
Buoyancy contrast within the ocean ($\Delta b = g \alpha \Delta T$)& 4 & m\,s$^{-2}$\\
Vertical viscosity ($A_z$) & 10$^1$ & m$^2$\,s$^{-1}$ \\
Vertical diffusivity ($k_z$) & $10^{-4}$ & m$^2$\,s$^{-1}$\\
Wind stress ($\tau$) & 0 & N\,m$^{-2}$\\
\hline
\multicolumn{3}{l}{}
\end{tabular}
\end{table}

Combining these three constraints, we are able to determine the characteristic scales for the magma ocean depth $D_{1t}$ as well as the horizontal current speed $U_{1t}$,
\begin{equation}\label{eq:thermal_scaling1}
D_{1t} \sim {(\frac{k_z A_z L^2} {\Delta b})}^{1/5}, \ U_{1t} \sim {(\frac{\Delta b^2 k_z^3 L}{A_z^2})}^{1/5}.
\end{equation}

Ocean circulation redistributes heat within the magma ocean basin. The resultant heating rate can drive the surface temperature away from radiative equilibrium. Following \citet{ferrari2011}, we estimate the ocean heat convergence ($H$),
\begin{equation}\label{eq:oht}
H = \frac{\partial}{\partial x}\int_{-D}^0 \rho c_p u \theta dz \sim \rho c_p \Delta T \frac{UD}{L},
\end{equation}
where $c_p$ represents heat capacity at constant pressure. By substituting the scaling laws of current speed $U_{1t}$ and ocean depth $D_{1t}$ (Equation (\ref{eq:thermal_scaling1})) into Equation (\ref{eq:oht}), we obtain the ocean heat convergence ($H_{1t}$)
\begin{equation}\label{eq:oht_scaling1_thermal}
H_{1t} = \frac{\rho c_p}{g \alpha} (\frac{k_z^4 \Delta b^4}{A_z L^2})^{1/5}.
\end{equation}

Under strong wind forcing, the main driving force of the ocean circulation shifts from thermal to wind forcing. Induced by winds and viscosity, there will be a boundary layer, in which wind forcing balances with vertical viscosity. 
Given the relatively large viscosity here, the boundary layer depth  
could be approximated to be comparable to the magma ocean depth. Thus, the momentum equation gives
\begin{equation}\label{eq:momentum_balance1_wind}
    \frac{1}{\rho_c} \frac{\partial \tau}{\partial z} \approx \frac{\partial}{\partial z}\left(A_z \frac{\partial u}{\partial z}\right),
\end{equation}
which scales as
\begin{equation}\label{eq:scaling_wind_viscosity}
    \frac{\tau}{\rho } \sim A_z \frac{U}{D}, 
\end{equation}
where $\tau$ represents wind stress. The momentum balance between the pressure gradient force and vertical viscosity still works in the interior ocean (Equation \eqref{eq:momentum_balance1}), despite the presence of wind stress. Thus, combining Equation (\ref{eq:scaling_momentum_balance1}) and Equation (\ref{eq:scaling_wind_viscosity}), the characteristic scales for the magma ocean depth $D_{1w}$ and horizontal velocity $U_{1w}$ under strong wind forcing are expressed as 
\begin{equation}\label{eq:wind_scaling1}
D_{1w} \sim (\frac{\tau L}{\rho \Delta b})^{1/2}, \ U_{1w} \sim (\frac{\tau^3 L}{\rho^3 A_z^2 \Delta b})^{1/2}.
\end{equation}
Substituting Equation (\ref{eq:wind_scaling1}) into Equation (\ref{eq:oht}), we obtain the ocean heat transport convergence under wind forcing ($H_{1w}$),
\begin{equation}\label{eq:oht_scaling1_wnd}
H_{1w} \sim \frac{c_p}{g \alpha}\frac{\tau^2}{\rho A_z}. 
\end{equation}

\begin{figure*}
    \centering
    \includegraphics[width=0.98\textwidth]{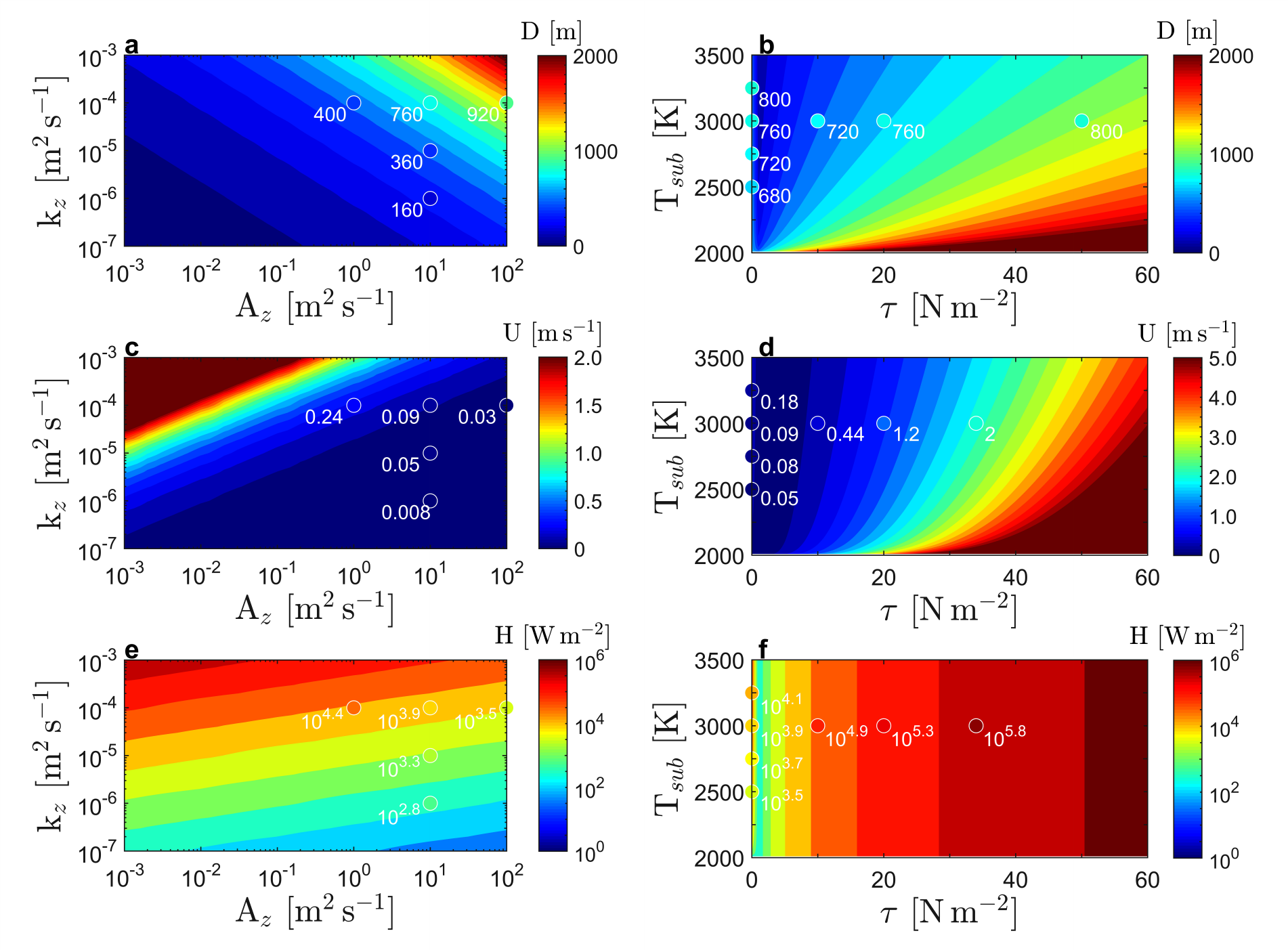}
    \caption{Scaling laws of the magma ocean depth ($D$; top), horizontal velocity ($U$; middle), and ocean heat transport convergence ($H$; bottom) in Regime I. Left panels: sensitivity of $D$, $U$, and $H$ to vertical viscosity ($A_z$) and vertical diffusivity ($k_z$) when dominated by thermal forcing. Right panels: sensitivity of $D$, $U$, and $H$ to wind stress ($\tau$) and substellar temperature ($T_{sub}$) when dominated by wind forcing. Note that in the right panels, results are still from the thermal forcing-dominated scaling when wind stress is zero. 
    Colors are results from scaling laws, and filled circles are results from numerical simulations. For numerical results, their exact values are indicated. }
    \label{fig:scaling_contour}
\end{figure*}

\begin{figure*}
    \centering
    \includegraphics[width=0.99\textwidth]{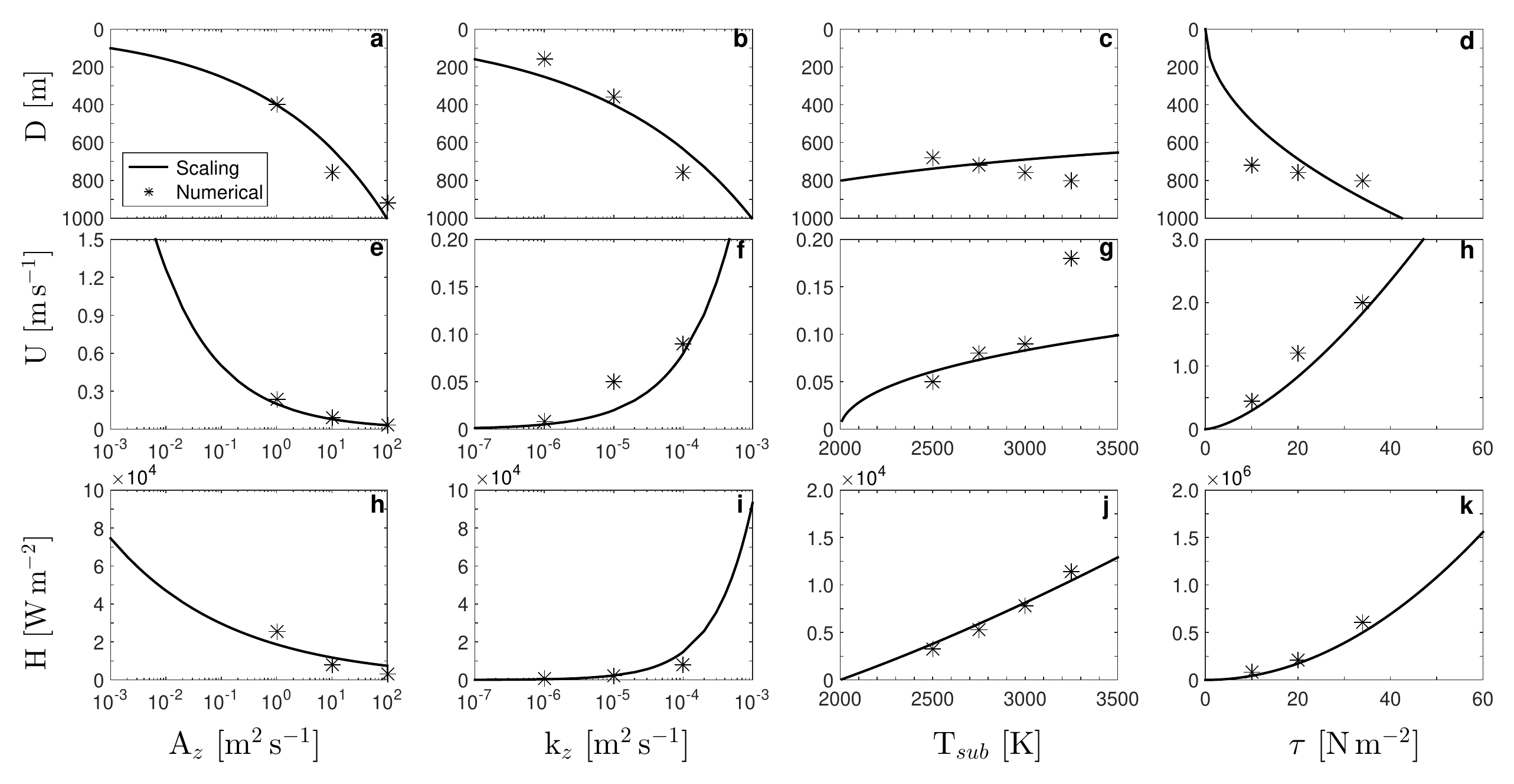}
    \caption{Comparison between results from scaling laws and from numerical simulations in Regime I. From left to right columns: sensitivity of the magma ocean depth $D$ (top), horizontal velocity $U$ (middle), and ocean heat transport convergence $H$ (bottom) to vertical viscosity ($A_z$), vertical diffusivity ($k_z$), substellar temperature ($T_{sub}$), and wind stress ($\tau$), respectively. The solid lines and scattered asterisks are results from scaling laws and numerical simulations, respectively. 
    By default, wind forcing is not included, except for the rightmost series of results.}
    \label{fig:scaling_lines}
\end{figure*}

To estimate and compare the magma ocean depth, horizontal current speed, and ocean heat transport convergence under varying parameters and forcings, we adopt the planetary parameters of Kepler-10b, including the planet radius ($a$), gravity ($g$), and the substellar temperature ($T_{sub}$) \citep{dumusque2014kepler}. The baseline values of planetary and oceanic parameters used in Figures \ref{fig:scaling_contour}--\ref{fig:scaling-regime3} are summarized in Table \ref{tab1}. By default, planetary rotation and wind stress are not included.
Notably, the temperature contrast within the magma ocean ($\Delta T$) is determined by the difference between the substellar temperture and the liquidus, while the horizontal scale of the magma ocean ($L$) is governed by the substellar temperature, the liquidus, and the planet radius. 

We start by discussing and validating the scaling results in a thermal-dominated system.
When the main driving factor of the circulation is thermal forcing, scaling laws indicate that magma ocean depth, horizontal velocity, and ocean heat transport convergence are determined by vertical diffusivity, surface temperature/buoyancy difference, and vertical viscosity (Equations (\ref{eq:thermal_scaling1}) \& (\ref{eq:oht_scaling1_thermal})).
As vertical viscosity/diffusivity increases, the influence of the surface forcings can reach deeper depth, increasing the magma ocean depth $D$. Then the variation of horizontal velocity differs when vertical viscosity and diffusivity changes. A higher vertical diffusivity deepens magma ocean depth, which then requests a faster flow speed to balance the same level of pressure gradient force (Equation (\ref{eq:scaling_momentum_balance1})). In contrast, a higher vertical viscosity results in weaker vertical upwelling speed $W$ as the diffusion pump needs to pump fluid from deeper depth (Equation (\ref{eq:scaling_advection_diffusion})). This in turn leads to a weaker horizontal flow $U$ under the constraint of mass continuity (Equation (\ref{eq:scaling_mass_continuity})). 
Due to the relatively limited change of ocean depth, the variation of heat transport convergence is dominated by the variation of horizontal velocity (Equation (\ref{eq:oht_scaling1_thermal})). Thus, heat transport convergence $H$ increases as vertical diffusivity increases and vertical viscosity decreases.

The scaling predictions for varying vertical viscosities are presented in Figure \ref{fig:scaling_contour}(a, c, e) and Figure \ref{fig:scaling_lines}(a, e, h). 
In order for the solutions to fall in the viscosity-dominant regime (Regime I), the vertical viscosity term needs to dominate the advection and Coriolis force terms (see Section \ref{sec:parameter_regime} for detailed discussion). This requires a relatively large vertical viscosity. Therefore, we only show scaling results for vertical viscosity ranging from 10$^{-3}$ to 10$^2$ m$^2$\,s$^{-1}$.

The dependence of the magma ocean depth, flow speed, and heat transport convergence on varying parameters predicted by scaling laws here, is confirmed by numerical simulations (scattered points of Figures \ref{fig:scaling_contour} \& \ref{fig:scaling_lines}).
The experiments presented here are taken from Part I of this series of papers. The model solves the zonal momentum, thermal, mass continuity, and sea surface height equations using finite difference method under two-dimensional(2D, x-z) Cartesian geometry. At the surface, the system is forced by thermal or wind forcing. The thermal forcing is represented by strong relaxation toward the radiative equilibrium temperature. The wind forcing is represented by prescribing the momentum flux injected into the upper surface of the ocean. At the bottom layer, vertical velocity is set to zero, and zonal velocity is strongly damped to guarantee an almost zero velocity. No-flux boundary condition is used for temperature there. Details of integration scheme and parameter choice can be found in Part I. 

As predicted, the simulated magma ocean depth $D$ deepens as vertical viscosity $A_z$ and vertical diffusivity $k_z$ increase. 
When $A_z$ increases from 10$^{0}$ to 10$^{2}$ m$^2$\,s$^{-1}$, $D$ deepens from about 400 to 920 m, accompanied by $U$ decreased from 0.24 to 0.03 m\,s$^{-1}$ (scattered points in Figure \ref{fig:scaling_contour}(a, c)). 
Meanwhile, heat transport convergence $H$ decreases from $2.5\times10^4$ to $3.2\times10^3$ W\,m$^{-2}$ (scattered points in Figure \ref{fig:scaling_contour}(e)).
When $k_z$ changes from 10$^{-6}$ to 10$^{-4}$ m$^2$\,s$^{-1}$, $D$ increases from 160 to 800 m, with $U$ intensified from 0.008 to 0.09 m\,s$^{-1}$, causing $H$ to increase from $5.8\times10^2$ to $7.8\times10^3$ W\,m$^{-2}$ (scattered points in Figure \ref{fig:scaling_contour}(a, c, e)). The results from numerical simulations well match that from scaling laws as can be seen in Figure \ref{fig:scaling_lines}(a, b, e, f, h, i). Since the heat convergence $H$ is 2-3 orders of magnitude smaller than the radiative flux, we do not expect the sea surface temperature to significantly deviate from the radiative equilibrium temperature, consistent with our assumptions.

The substellar temperatures of lava planets vary significantly \citep{leger2011extreme,castan2011atmospheres,batalha2011kepler,demory2011detection,nguyen2020modelling}. Here, we change the substellar temperature from 2000 to 3500 K, and the corresponding scaling predictions are presented in Figure \ref{fig:scaling_contour}(b, d, f) and Figure \ref{fig:scaling_lines}(c, g, j). Given that the liquidus is assumed to be 2000 K \citep{monteux2016cooling}, the correspondent temperature difference ranges from 0 to 1500 K.
As the substellar temperature increases, two things happen simultaneously. First, the surface temperature difference will increase, and that will result in a stronger pressure gradient force. Consequently, both horizontal and vertical velocities increase, which means the gravitational potential energy is consumed and the stratification is enhanced at a faster rate. Following the advective-diffusive balance, a fixed vertical diffusivity can only pump dense fluid up faster if the pumping depth if shallower (Equation (\ref{eq:scaling_advection_diffusion})).
Second, the width of magma ocean also increases, and that will lead to a deeper magma ocean under the constraint of mass continuity (Equation (\ref{eq:scaling_mass_continuity})).
The two effects cancel each other, making magma ocean depth insensitive to the change of substellar point temperature.
The change of heat transport convergence is dominated by the variation of current speed, which intensifies as the substellar temperature increases. 

Numerical results under varying substellar temperatures $T_{sub}$ show that as $T_{sub}$ rises from 2500 to 3250 K, the magma ocean depth $D$ increases from 680 to 800 m, with horizontal velocity $U$ increased from 0.05 to 0.18 m\,s$^{-1}$ (scattered points of Figure \ref{fig:scaling_contour}(b, d)). 
As a result, heat transport convergence increases from $3.3\times10^3$ to $1.2\times10^4$ W\,m$^{-2}$ (scattered points of Figure \ref{fig:scaling_contour}(f)).
Simulation results indicate that the dependence of magma ocean depth on substellar temperature is relatively weak, consistent with predictions from scaling laws. However, for unclear reasons, the two trends are opposite (Figure \ref{fig:scaling_lines}(c)).


We then discuss the scaling results in a wind-dominated system and verify the scaling using numerical experiments (Equations (\ref{eq:wind_scaling1}) \& (\ref{eq:oht_scaling1_wnd})) .
Here, scaling predictions are presented in Figure \ref{fig:scaling_contour}(b, d, e) and Figure \ref{fig:scaling_lines}(d, h, k) for wind stress between 0 and 60 N\,m$^{-2}$.
As wind stress increases, scaling laws predict that magma ocean depth,  horizontal velocity, and heat transport convergence should increase. In particular, the wind forcing on Kepler-10b might reach a maximum value of 100 N\,m$^{-2}$, and it does dominate the thermal forcing as the major driver of the ocean circulation. 

We conduct numerical simulations with wind stress $\tau$ set to 10, 20, and 34 N\,m$^{-2}$, and the results are shown by scattered points in Figure \ref{fig:scaling_contour}(b, d, f). When $\tau$ increases from 10 to 34 N\,m$^{-2}$, magma ocean depth $D$ increases from 720 to 800 m, accompanied by horizontal velocity $U$ intensified from 0.44 to 2.0 m\,s$^{-1}$. 
The increase of both $D$ and $U$ results in the increase of $H$, which changes from $7.8\times10^4$ to $6.1\times10^5$ W\,m$^{-2}$. 
Results from scalings and simulations are consistent (Figure \ref{fig:scaling_lines}(d, h, k)). 


\subsection{Regime II: non-rotating inviscid limit regime}\label{scaling_invis}

Viscosity of fully molten silicates might be decreased to 10$^{-4}$ or $10^{-5}$ m$^2$\,s$^{-1}$ \citep{sun2020physical,zhang2022internal}. Thus, solutions in Regime I might not work when an extremely weak viscosity is considered.
Here, we present scaling laws to the horizontal current speed, magma ocean depth, and ocean heat transport convergence under non-rotating inviscid limit regime. Similar to Regime I, Coriolis force is neglected. 
Under inviscid limit, the pressure gradient force is balanced by the nonlinear advection in the momentum budget (blue brackets in Figure \ref{fig:equation}(a)). Then the dominant balance of momentum equation is
\begin{equation}\label{eq:momentum_balance2}
    \frac{1}{\rho_c} \frac{\partial p}{\partial x} \approx u\frac{\partial u}{\partial x},
\end{equation}

which scales as,
\begin{equation}\label{eq:scaling_momentum_balance2}
\frac{\Delta b}{L} D \sim \frac{U^2}{L}.  
\end{equation}
Similar to that in Regime I, mass continuity needs to be satisfied within the magma ocean, and the buoyancy is in balance between upwelling and diffusion. 
Employing Equations (\ref{eq:scaling_mass_continuity}), (\ref{eq:scaling_advection_diffusion}), and (\ref{eq:scaling_momentum_balance2}), magma ocean depth ($D_{2t}$) and horizontal current speed ($U_{2t}$) scale as,
\begin{equation}\label{eq:thermal_scaling_regime2}
D_{2t} \sim {(\frac{k_z^2 L^2} {\Delta b})}^{1/5}, \ U_{2t} \sim {(\Delta b^2 k_z L)}^{1/5}.
\end{equation} 
Substituting Equation (\ref{eq:thermal_scaling_regime2}) into Equation (\ref{eq:oht}), we obtain the ocean heat transport convergence ($H_{2t}$)
\begin{equation}\label{eq:oht_scaling2_thermal}
H_{2t} \sim \frac{\rho c_p}{g \alpha} (\frac{k_z^3 \Delta b^6}{L^2})^{1/5}.
\end{equation}

In the presence of wind forcing, the effect of wind forcing depends on the specific value of the vertical viscosity. If viscosity of the magma surface is entirely zero, the magma ocean will not have any influence on the atmospheric momentum. In other words, the magma ocean will not be affected by the atmospheric winds.
But if the viscosity is not exactly zero but a small number, wind stress can influence the magma ocean potentially. Here, we estimate the impacts assuming a small viscosity.

\begin{figure*}
    \centering
    \includegraphics[width=0.95\textwidth]{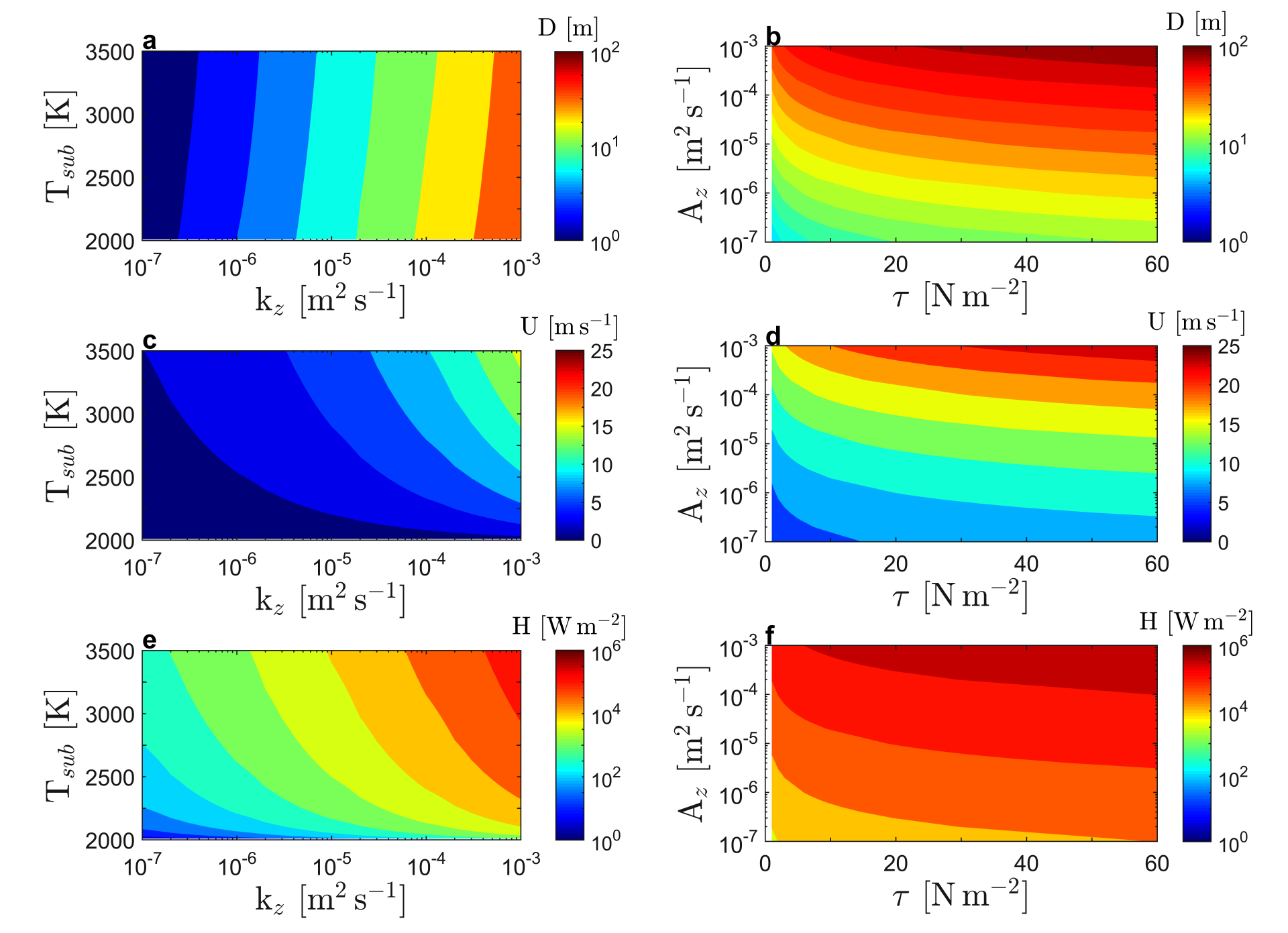}
    \caption{Scaling laws of the magma ocean depth ($D$; top), horizontal velocity ($U$; middle), and ocean heat transport convergence ($H$; bottom) in Regime II. Left panels: sensitivity of $D$, $U$, and $H$ to vertical diffusivity ($k_z$) and substellar temperature ($T_{sub}$) when dominated by thermal forcing. Right panels: sensitivity of $D$, $U$, and $H$ to wind stress ($\tau$) and vertical viscosity ($A_z$) when dominated by wind forcing. }
    \label{fig:scaling-regime2}
\end{figure*}

Given the smallness of viscosity, only a very thin boundary layer near the surface will be directly influenced by wind stress. There, the dominant momentum balance is between the vertical viscosity and nonlinear advection, and the vertical viscosity scales with wind stress
\begin{equation}\label{eq:momentum_balance2_wind}
   u \frac{\partial u}{\partial x} \approx \frac{\partial}{\partial z}\left(A_z \frac{\partial u}{\partial z}\right)\sim \frac{1}{\rho_c}\frac{\partial \tau}{\partial z}.
\end{equation}
This leads to the following scale relationships
\begin{equation}\label{eq:scaling_momentum_balance2_wind}
\frac{U_v^2}{L}\sim A_z \frac{U_v}{D_v^2} \sim \frac{\tau}{\rho D_v},
\end{equation}
from which $U_v$ and $D_v$ are expressed as
\begin{equation}\label{eq:scaling_Uv2}
U_v \sim (\frac{\tau^2 L}{\rho^2 A_z})^{1/3}, \ D_v \sim (\frac{\rho A_z^2 L}{\tau})^{1/3},
\end{equation}
where $U_v$ and $D_v$ represent the typical scales of horizontal velocity in the boundary layer and the depth of the boundary layer, respectively.

Then the interior vertical velocity $W$ induced by the horizontal convergence/divergence of mass transport in the boundary layer should follow
\begin{equation}\label{eq:scaling_Wv2}
W \sim \frac{U_v D_v}{L} \sim (\frac{\tau A_z}{\rho L})^{1/3}.
\end{equation}
Replacing Equation (\ref{eq:scaling_advection_diffusion}) and employing Equations (\ref{eq:scaling_mass_continuity}), (\ref{eq:scaling_momentum_balance2}), and (\ref{eq:scaling_Wv2}), the scales for the magma ocean depth ($D_{2w}$) and horizontal velocity ($U_{2w}$) are,
\begin{equation}\label{eq:scaling_wind2}
D_{2w} \sim (\frac{\tau^2 A_z^2 L^4}{\rho^2 \Delta b^3})^{1/9}, \ U_{2w} \sim (\frac{\Delta b^3 \tau A_z L^2}{\rho})^{1/9}.
\end{equation}
Similarly,  substituting Equation (\ref{eq:scaling_wind2}) to Equation (\ref{eq:oht}), the scaling for ocean heat transport convergence ($H_{2w}$) is written as
\begin{equation}\label{eq:oht_scaling2_wnd}
H_{2w} \sim \frac{g}{g \alpha} (\rho^2 \Delta b^3 \frac{\tau A_z}{L})^{1/3}.
\end{equation}


Results from the scaling laws in Regime II are presented in Figure \ref{fig:scaling-regime2}. Due to the weak viscosity/diffusivity, the magma ocean depth will not exceed 100 m across the parameter range we consider here, much thinner than that in Regime I.

When dominated by thermal forcing, magma ocean depth, horizontal velocity, and heat transport convergence are mainly determined by vertical diffusivity and surface temperature difference (Equations (\ref{eq:thermal_scaling_regime2}) \& (\ref{eq:oht_scaling2_thermal})). Similar to Regime I, as vertical diffusivity increases, magma ocean deepens following the advection-diffusion balance (Equation (\ref{eq:scaling_advection_diffusion})), accompanied by increased horizontal velocity and heat transport convergence (Figure \ref{fig:scaling-regime2}(a, c, e)). Varying substellar temperatures lead to changes in the surface temperature difference, i.e., the pressure gradient force. Thus, horizontal velocity increases under larger temperature difference (higher substellar temperature), 
leading to stronger heat transport convergence. Meanwhile, magma ocean becomes shallower due to stronger vertical velocity (Figure \ref{fig:scaling-regime2}(a, c, e)). 

When dominated by wind forcing, magma ocean depth, horizontal velocity, and heat transport convergence are affected by the amplitude of wind stress and vertical viscosity (Equations (\ref{eq:scaling_wind2}) \& (\ref{eq:oht_scaling2_wnd})). 
Results with vertical viscosity ranging from 10$^{-7}$ to 10$^{-3}$ m$^2$\,s$^{-1}$ are shown in Figure \ref{fig:scaling-regime2}(b, d, f), in which vertical viscosity discussed is much smaller than that in Regime I. As vertical viscosity decreases, wind-induced boundary layer is shallower, accompanied by weaker horizontal velocity in the boundary layer, which will then lead to weakening of vertical velocity in the interior (Equations \eqref{eq:scaling_Uv2} and (\ref{eq:scaling_Wv2})). Magma ocean depth here, determined by the horizontal and vertical advections, becomes shallower.  
The interior horizontal velocity is determined by the balance between the pressure gradient force and nonlinear advection (Equation (\ref{eq:scaling_momentum_balance2})). Thus, shallower magma ocean corresponds to smaller pressure gradient, decreasing the horizontal velocity. 
The decreases of both ocean depth and current speed result in the decrease of heat transport convergence (Equation (\ref{eq:oht_scaling2_wnd})).

It should be noted that the value of vertical viscosity in this regime should be small than certain values to satisfy the momentum balance here. As we introduce in the beginning of this section that the momentum equation is in balance between the pressure gradient force and advection below the boundary layer. Thus, vertical viscosity term is not allowed to be larger than advection. The specific values of vertical viscosity in Regime II will be discussed in Section \ref{sec:parameter_regime} in detail.

As wind stress increases, magma ocean depth reaches deeper and horizontal velocity becomes larger, leading to intensified heat transport convergence (Figure \ref{fig:scaling-regime2}(b, d, f)). The increase in wind stress directly makes the horizontal divergence/convergence of mass transport larger, then intensifying the vertical velocity. 
Thus, magma ocean deepens with stronger vertical advection. Deeper magma ocean increases the vertical integral of density contrast, i.e., the pressure gradient force. Thus, the interior horizontal velocity becomes larger under stronger wind stress.


\subsection{Regime III: rotation-dominant regime}\label{scaling_3D}

In Sections \ref{scaling_2D} and \ref{scaling_invis}, we present scaling laws assuming rotation effect is negligible. Here, we present scaling laws for the horizontal current speed, magma ocean depth, and ocean heat transport convergence when rotation plays a dominant role.
In this regime, Coriolis force balances the pressure gradient force (green brackets in Figure \ref{fig:equation}(a)). This so-called geostrophic balance regime has been extensively studied due to its application to Earth's oceans. Scaling laws derived for Earth's oceans can also be applied to thermally-forced lava oceans as done in \citet{kite2016atmosphere}. Here, we will briefly review the derivation. Interested readers can refer to textbooks such as \citet{vallis2017atmospheric} and \citet{vallis2019essentials} for more information. 

The derivation utilizes three relationships. Firstly, by taking the vertical curl of the horizontal momentum equation and using the mass continuity, we can derive the linear vorticity equation 
\begin{equation}\label{eq:3D_vort}
   \beta v=f \frac{\partial w}{\partial z},
\end{equation}
where $\beta=2\Omega \cos\phi/a$ denotes the gradient of the Coriolis coefficient $f= 2 \Omega \sin\phi$ along the meridional direction, $\Omega$ denotes the planetary rotation rate, $\phi$ denotes latitude, and $a$ denotes planet radius. In the linear vorticity equation, the LHS represents the advection of planetary vorticity, and the RHS represents the generation of vorticity by vortex stretching. 

Secondly, by taking the vertical derivative of the momentum equation (geostrophic balance relation) and using the hydrostatic balance, we derive the thermal wind balance,
\begin{equation}\label{eq:3D_thermalwind}
f\frac{\partial\mathbf{u}}{\partial z}=\mathbf{k}\times\nabla b,
\end{equation}
where $\mathbf{u}$ denotes horizontal velocity and $\mathbf{k}$ denotes the unit vector pointing upward. 
The last set of relationship has to do with the driving force of the circulation. 
When wind forcing is weak, the dominant driver of the circulation is diffusion.
Similar to previous cases, the heat/buoyancy will be in advection-diffusion balance (Equation (\ref{eq:advective_diffusive_balance})).

We can then write down the corresponding scales for Equations \eqref{eq:3D_vort}, \eqref{eq:3D_thermalwind} and \eqref{eq:advective_diffusive_balance} as
\begin{subequations}\label{eq:3D_equation_thermal}
\begin{equation}\label{eq:3D_equation1_thermal}
\beta U \sim f \frac{W}{D},
\end{equation}
\begin{equation}\label{eq:3D_equation2_thermal}
f\frac{U}{D} \sim \frac{\Delta b}{L},
\end{equation}
\begin{equation}\label{eq:3D_equation3_thermal}
\quad D \sim \frac{k_z}{W}.
\end{equation}
\end{subequations}
It should be noted that we assume $U$ $\sim$ $V$ in the scaling analysis presented here. This assumption is valid for tidally locked lava planets, where the latitudinal and longitudinal extension of  magma ocean are comparable. In other words, although $U$ is not directly affected by the beta effect, it also obeys the Svedrup balance in accordance with $V$.
From Equation (\ref{eq:3D_equation_thermal}), the characteristic scales for the magma ocean depth ($D_{3t}$) and horizontal velocity ($U_{3t}$) can be solved,
\begin{equation}\label{eq:scaling_thermal3}
D_{3t} \sim (\frac{k_z f^2 L} {\beta \Delta b})^{1/3}, \  U_{3t} \sim (\frac{k_z \Delta b^2} {\beta f L^2})^{1/3}.
\end{equation}
Combining Equations (\ref{eq:oht}) and (\ref{eq:scaling_thermal3}), the scale for ocean heat transport convergence ($H_{3t}$) is expressed as,
\begin{equation}\label{eq:oht_scaling3_thermal}
H_{3t} \sim \frac{\rho c_p}{g \alpha} (\frac{fk_z^2\Delta b^4}{\beta^2L^4})^{1/3}.
\end{equation}

\begin{figure*}
    \centering
    \includegraphics[width=0.95\textwidth]{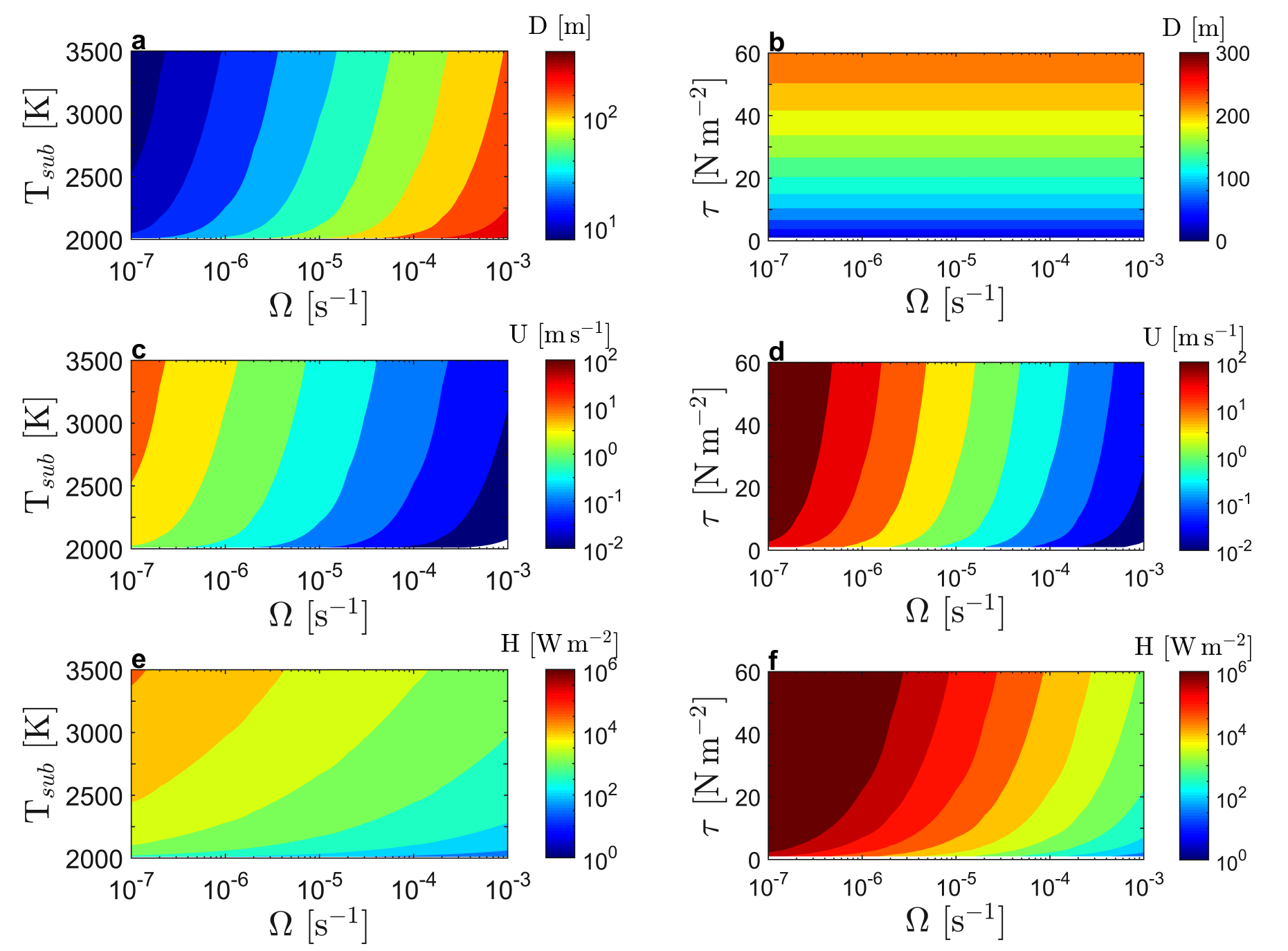}
    \caption{Scaling laws of the magma ocean depth ($D$; top), horizontal velocity ($U$; middle), and ocean heat transport convergence ($H$; bottom) in Regime III. Left panels: sensitivity of $D$, $U$, and $H$ to planetary rotation rate ($\Omega$) and substellar temperature ($T_{sub}$) when dominated by thermal forcing. Right panels: sensitivity of $D$, $U$, and $H$ to planetary rotation rate ($\Omega$) and wind stress ($\tau$) when dominated by wind forcing.}
    \label{fig:scaling-regime3}
\end{figure*}


When wind forcing is the main driving force of the circulation, we need to consider the interior vertical motions induced by the curl of wind stress, a phenomenon known as Ekman pumping. The vertical velocity at the base of the Ekman layer $W_E$ follows \citep{1996Pedosky,2008Stewart,vallis2019essentials}, 
\begin{equation}\label{eq:Ekman_pumping_wind}
W_E = - \frac{1}{\rho}curl_z(\frac{\tau}{f})\sim \frac{\tau}{\rho L f},
\end{equation}
where $\tau$ is the wind stress and $curl_z$ represents the vertical component of curl.
Replacing Equation (\ref{eq:3D_equation3_thermal}) with Equation \eqref{eq:Ekman_pumping_wind}, we instead get the following scaling laws for the magma ocean depth ($D_{3w}$) and horizontal velocity ($U_{3w}$), 
\begin{equation}\label{eq:scaling_wind3}
D_{3w} \sim (\frac{\tau f}{\rho \beta \Delta b})^{1/2}, \ U_{3w} \sim (\frac{\tau \Delta b}{\rho f\beta L^2})^{1/2}.
\end{equation}
Substitute Equation (\ref{eq:scaling_wind3}) into Equation (\ref{eq:oht}), we obtain the characteristic scale for ocean heat transport convergence ($H_{3w}$),
\begin{equation}\label{eq:oht_scaling3_wnd}
H_{3w} \sim \frac{c_p}{g \alpha}\frac{\tau \Delta b}{\beta L^2}.
\end{equation}
The scaling laws for magma ocean depth ($D$), horizontal current speed ($U$), and ocean heat transport convergence ($H$) under varying dominant forcings and regimes are summarized in Table \ref{tab:scaling_sumamry}. 



\begin{table}
\caption{Summary of Scaling Laws under Varying Dominant Forcings and Regimes
\label{tab:scaling_sumamry}}
\centering
\begin{threeparttable}
\begin{tabular}{llccc}
\hline
Forcing   &  Regime    &  Ocean depth (D)  &  Current speed (U) & Heat transport convergence (H) \\ \hline
Thermal         & I    & $(\frac{k_z A_z L^2}{\Delta b})^{1/5}$ &  $(\frac{\Delta b^2 k_z^3 L}{A_z^2})^{1/5}$ & $\frac{\rho c_p}{g\alpha} (\frac{k_z^4 \Delta b^6 }{A_z L^2})^{1/5}$ \\ 
                & II   & $(\frac{k_z^2 L^2}{\Delta b})^{1/5}$ & $(\Delta b^2 k_z L)^{1/5}$ & $\frac{\rho c_p}{g\alpha} (\frac{k_z^3 \Delta b^6 }{L^2})^{1/5}$ \\ 
                & III  & $(\frac{k_z f^2 L}{\beta \Delta b})^{1/3}$ & $(\frac{k_z \Delta b^2}{\beta f L^2})^{1/3}$ & $\frac{\rho c_p}{g\alpha} (\frac{f k_z^2 \Delta b^4 }{\beta^2 L^4})^{1/3}$ \\ \cline{1-5}
Wind            & I    & $(\frac{\tau L}{\rho \Delta b})^{1/2}$ &  $(\frac{\tau^3 L}{\rho^3 A_z^2 \Delta b})^{1/2}$ & $\frac{c_p}{g\alpha} (\frac{\tau^2}{\rho A_z})$ \\ 
                & II   & $(\frac{\tau^2 A_z^2 L^4}{\rho^2 \Delta b^3})^{1/9}$ & $(\frac{\tau A_z \Delta b^3 L^2}{\rho })^{1/9}$ & $\frac{c_p}{g\alpha} (\frac{\tau A_z}{L}\rho^2 \Delta b^3)^{1/3}$ \\ 
                & III  & $(\frac{\tau f}{\rho \beta \Delta b})^{1/2}$ & $(\frac{\tau \Delta b}{\rho f \beta L^2})^{1/2}$ & $\frac{c_p}{g\alpha} (\frac{\tau \Delta b}{\beta L^2})$ \\  \cline{1-5}              
\end{tabular}
\begin{tablenotes}
\footnotesize
\item  In the scaling laws, $k_z$ is vertical diffusivity, $A_z$ is vertical viscosity, $L$ is the horizontal scale of the magma ocean, $\Delta b$ is horizontal buoyancy contrast, $\tau$ is wind stress, $f$ is Coriolis parameter, $\beta$ is the meridional gradient of Coriolis parameter, $\rho$ is ocean density, $g$ is gravity, $\alpha$ is thermal expansion coefficient, and $c_p$ is heat capacity at constant pressure. 
\end{tablenotes}
\end{threeparttable}
\end{table}

When wind forcing is absent or weak, Equations (\ref{eq:scaling_thermal3}) and (\ref{eq:oht_scaling3_thermal}) suggest that the magma ocean depth, horizontal velocity, and ocean heat transport convergence are controlled by vertical diffusivity, planetary rotation rate, and surface temperature difference. The effects of vertical diffusivity have been illustrated many times previously. In Figure \ref{fig:scaling-regime3}(a, c, e), results under varying rotation rates and substellar temperatures are presented. As rotation rate increases, horizontal velocity reduces under the geostrophic balance constraint. Thus, vertical velocity decreases from mass continuity, leading to a deeper magma ocean. As the substellar temperature increases, the pressure gradient force is larger, resulting in stronger horizontal and vertical velocities. Meanwhile, magma ocean depth, determined by advection-diffusion balance, is decreased. 
Dominated by the variation of current speed, ocean heat transport convergence decreases with increasing rotation rate, while increases as substellar temperature becomes higher (Equation (\ref{eq:oht_scaling3_thermal})).

In the presence of strong wind forcing, the wind stress, planetary rotation rate, and substellar temperature could exert influences on the magma ocean depth, horizontal velocity, and heat transport convergence (Equations (\ref{eq:scaling_wind3}) \& (\ref{eq:oht_scaling3_wnd})).
Here we present results under varying wind stresses and planetary rotation rates (Figure \ref{fig:scaling-regime3}(b, d, f)).  
As rotation rate increases, magma ocean depth remains unchanged, accompanied by a smaller horizontal velocity. According to the geostrophic balance and Ekman pumping relationship, both horizontal and vertical velocities will decrease with rotation rate at a same rate (Equations \eqref{eq:3D_equation2_thermal} and \eqref{eq:Ekman_pumping_wind}). Thus, the effect of varying rotation rates on magma ocean depth could vanish (Equation \eqref{eq:3D_equation1_thermal}). 
Weaker ocean circulation results in decreased ocean heat transport convergence (Equation (\ref{eq:oht_scaling3_wnd})).

The influences of varying amplitudes of wind stress are presented in Figure \ref{fig:scaling-regime3}(b, d, f). As wind stress increases, both ocean depth and horizontal velocity increases. This is because Ekman pumping (subduction) strengthens under larger wind stress, which deepens the magma ocean. With deeper ocean, the vertical integral of buoyancy contrast increases, leading to larger pressure gradient force and then stronger horizontal velocity. Proportional to current speed and ocean depth, ocean heat transport convergence increases as wind stress becomes stronger.


\section{Parameter regimes for the scalings}\label{sec:parameter_regime}
We have presented different scaling laws in different dynamical regimes in Section \ref{sec:theoretical}, based on the balance between the horizontal pressure gradient force and varying terms.
In this section, we attempt to discuss the parameter space where each of the scaling laws is applicable. Especially, the parameter regime is discussed in two parts: thermal forcing-dominated and wind forcing-dominated systems.  

\subsection{Thermal forcing-dominated system}\label{parameter_thermal}

In Regime I, the pressure gradient force is balanced by vertical viscosity, i.e., the vertical viscosity term is greater than both the advection and Coriolis force. Thus, the relationship between the vertical viscosity, advection, and Coriolis force can be expressed as,
\begin{equation}\label{eq:parameter_regime1}
A_z \frac{U}{D^2} > max\{\frac{U^2}{L}, fU\}.
\end{equation}
Substituting $U_{1t}$ and $D_{1t}$ from Equation (\ref{eq:thermal_scaling1}) into Equation (\ref{eq:parameter_regime1}), we derive the relationship between varying parameters, 
\begin{equation}\label{eq:parameter_regime1-2}
A_z > max\{k_z, \Omega^{\frac{5}{3}}(\frac{k_z L^2}{\Delta b})^{\frac{2}{3}}\}.
\end{equation}
Thus, the magma ocean will be in Regime I when the vertical viscosity coefficient is larger than certain values. 

Regime II occurs when the nonlinear advection term is larger than both the vertical viscosity and Coriolis force, i.e.,
\begin{equation}\label{eq:parameter_regime2}
\frac{U^2}{L} > max\{A_z \frac{U}{D^2}, fU\}.
\end{equation}
Substituting the scaling laws for the horizontal current speed and ocean depth in Regime II, i.e., $U_{2t}$ and $D_{2t}$ from Equation (\ref{eq:thermal_scaling_regime2}), into Equation (\ref{eq:parameter_regime2}), we obtain that 
\begin{equation}\label{eq:parameter_regime2-2}
A_z < k_z, \Omega < (\frac{\Delta b^2 k_z}{L^4})^{\frac{1}{5}}.
\end{equation}
Thus, the magma ocean will be in Regime II when both the vertical viscosity coefficient and planetary rotation rate are smaller than certain values.

Magma ocean is in Regime III when the Coriolis force is greater than both the vertical viscosity and nonlinear advection, which is expressed as,
\begin{equation}\label{eq:parameter_regime3}
fU > max\{A_z \frac{U}{D^2}, \frac{U^2}{L}\}.
\end{equation}
It is worth noting that Equation (\ref{eq:parameter_regime3}) can also be represented as both the vertical Ekman Number ($E_z = \frac{A_z}{f D^2}$) and the Rossby Number ($R_o = \frac{U}{fL}$) are smaller than one \citep{2008Stewart,vallis2019essentials}. Employing the scaling laws for the horizontal current speed and ocean depth in Regime III, i.e., $U_{3t}$ and $D_{3t}$ from Equation (\ref{eq:scaling_thermal3}), Equation (\ref{eq:parameter_regime3}) becomes
\begin{equation}\label{eq:parameter_regime3-2}
\Omega > max\{ A_z^{\frac{3}{5}} (\frac{\Delta b}{k_z L a})^{\frac{2}{5}}, (\frac{k_z \Delta b^2 a}{L^5})^{\frac{1}{5}}\}.
\end{equation}

\begin{figure*}
    \centering
    \includegraphics[width=0.98\textwidth]{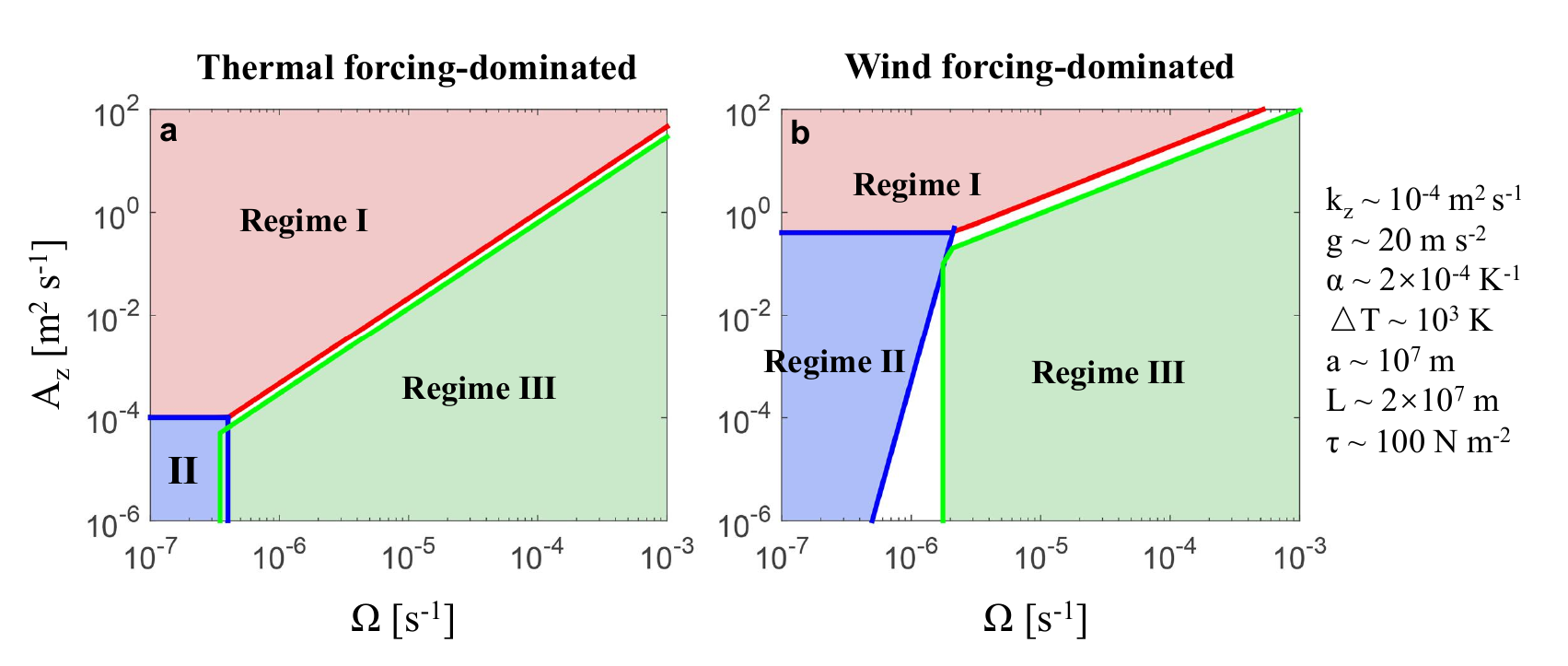}
    \caption{Regime diagram in ($\Omega$, $A_z$)-space in thermal forcing-dominated (a) and wind forcing-dominated (b) systems. The regions in red, blue, and green colors correspond to the parameter space in Regimes I (viscosity-dominant), II (advection-dominant), and III (rotation-dominant), respectively. 
    The parameters used to obtain this diagram, including vertical diffusivity ($k_z$), gravity ($g$), thermal expansion coefficient ($\alpha$), temperature difference ($\Delta T$), planet radius ($a$), horizontal range of the magma ocean ($L$), and wind stress ($\tau$), are listed on the right side of panel (b). Note that wind stress is not included ($\tau$ = 0) in the thermal forcing-dominated case. 
  Due to the simplicity of these scaling laws, there are some overlap between different regimes, and blank areas are not resolved by these scalings.}
    \label{fig:parameter-regime}
\end{figure*}

Equations (\ref{eq:parameter_regime1-2}), (\ref{eq:parameter_regime2-2}), and (\ref{eq:parameter_regime3-2}) present the criteria for the three scaling laws to hold when wind forcing is negligible compared to thermal forcing. 
To give an example, we assume a super-Earth type of configuration: $k_z \sim 10^{-4}$ m$^2$\,s$^{-1}$, $g \sim$ 20 m\,s$^{-2}$, $\alpha \sim 2\times10^{-4}$ K$^{-1}$, $\Delta T \sim 10^3$ K, $\Delta b \sim$ 4 m\,s$^{-2}$, $L \sim 2\times10^7$ m, and $a \sim 10^7$ m. The criteria for the three regimes follow
\begin{equation}\label{eq:parameter_regime}
\begin{aligned}
& \rm {Regime\ I:} \, A_z > max\{10^{-4}, 5\cdot10^6 \times \Omega^{\frac{5}{3}}\},\\
& \rm {Regime\ II:} \, A_z < 10^{-4} , \Omega < 4\times10^{-7},\\
& \rm {Regime\ III:} \, \Omega > max\{10^{-4}\times A_z^{\frac{3}{5}}, 4\times10^{-7} \}.
\end{aligned}
\end{equation}
In Figure \ref{fig:parameter-regime}(a), we mark the three regimes using different colors in the $(\Omega, A_z)$ parameter space.  
Since the viscosity of the molten rock (i.e., the magma ocean) is usually around 10$^{-5}$ or 10$^{-4}$~m$^2$\,s$^{-1}$ \citep{zhang2022internal}, and the orbital period of a typical tidally locked lava planet (such as Kepler-10b, CoRot-7b, 55 Cnc e, and K2-141b) is around one Earth day \citep{leger2011extreme,batalha2011kepler,bourrier201855,malavolta2018ultra,brinkman2023toi}, the rotation-dominant Regime III is likely to be the most relevant.


With a 10$^{-4}$~m$^2$\,s$^{-1}$ viscosity, Regime I and II will only be relevant if the orbital period is longer than $\sim$180 Earth days. This may be achieved on slowly-rotating exoplanets or moons, whose magma ocean is created by giant impacts \citep{elkins2012magma}.
It should be noted that although the molecular viscosity/diffusivity is generally weak, when turbulence is present, the eddy viscosity/diffusivity might be much greater than $10^{-4}$ m$^2$\,s$^{-1}$. For example, in Earth's oceans, the vertical eddy viscosity/diffusivity could reach $10^{-3}$--$10^{-1}$ m$^2$\,s$^{-1}$, in places with strong topography, winds, baroclinic eddies \citep{luo2023study,sentchev2023estimation}. Besides these energy sources, lava worlds also subject to strong tidal forcing, and that can further trigger internal gravity waves and turbulent mixing in the ocean \citep{bell1975a, smithyoung2001}. 
With a larger eddy viscosity of $10^{-2}$ m$^2$\,s$^{-1}$, the magma ocean could switch to Regime I with an orbital period beyond $\sim$10 Earth days.


\subsection{Wind forcing-dominated system}\label{parameter_wind}

Following the same approach, in Regime I, the vertical viscosity term should be greater than both the nonlinear advection and Coriolis force, i.e., Equation (\ref{eq:parameter_regime1}). Employing the scaling laws for horizontal current speed ($U_{1w}$) and ocean depth ($D_{1w}$) in Regime I from Equation (\ref{eq:wind_scaling1}), we get the following criterion 
\begin{equation}\label{eq:parameter_wind-regime1}
A_z > max\{(\frac{\tau_x^5 L}{\rho^5 \Delta b^3})^{\frac{1}{4}}, \ \Omega \frac{\tau_x L}{\rho \Delta b} \}.
\end{equation}
In Regime II, the nonlinear advection should be larger than both the vertical viscosity and Coriolis force. Substituting the scaling laws for $U_{2w}$ and $D_{2w}$ from Equation \eqref{eq:scaling_wind2} into Equation (\ref{eq:parameter_regime2}), we obtain that
\begin{equation}\label{eq:parameter_wind-regime2}
A_z < (\frac{\tau_x^5 L}{\rho^5 \Delta b^3})^{\frac{1}{4}}, \ \Omega < A_z^{\frac{1}{9}} (\frac{\Delta b^3 \tau_x}{\rho L^7})^{1/9}.
\end{equation}
In Regime III, the Coriolis force dominates over both the vertical viscosity and nonlinear advection, i.e., Equation (\ref{eq:parameter_regime3}).
Adopting the scaling laws for $D_{3w}$ and $U_{3w}$ from Equation (\ref{eq:scaling_wind3}), we obtain that
\begin{equation}\label{eq:parameter_wind-regime3}
\Omega > max\{A_z \frac{\rho \Delta b}{ \tau a}, \ (\frac{\tau \Delta b }{\rho  L^4})^{\frac{1}{4}} \}.
\end{equation}

We revisit the example mentioned in section~\ref{parameter_thermal} assuming $\tau \sim$ 100 N\,m$^{-2}$ \citep{kang2021escaping}, 
and criteria (\ref{eq:parameter_wind-regime1})--(\ref{eq:parameter_wind-regime3}) can be written as,
\begin{equation}\label{eq:parameter_regime_wind}
\begin{aligned}
& \rm {Regime\ I:} \, A_z > max\{0.4, \ 2\cdot10^{5} \times \Omega \},\\
& \rm {Regime\ II:} \, A_z < 0.4, \ \Omega < 2\cdot10^{-6}\times A_z^{1/9}\\
& \rm {Regime\ III:} \, \Omega > max\{10^{-5}\times A_z, \ 2\times10^{-6} \}.
\end{aligned}
\end{equation}

The regime diagram for a wind-dominated system is presented in Figure \ref{fig:parameter-regime}(b). Given the typical vertical viscosity of the magma ocean ($A_z \sim 10^{-4}$ m$^2$\,s$^{-1}$) and the rotation rate of tidally locked lava planets ($\Omega \sim 10^{-4}$ m$^{-1}$), the ocean circulation will be most likely in the rotation-dominant Regime III. The dominant role of rotation in the ocean circulation works both in thermal forcing-dominated and wind forcing-dominated systems. 
With a 10$^{-4}$ m$^2$\,s$^{-1}$ viscosity, the magma ocean could shift to Regime II when the orbital period is beyond $\sim$100 Earth days. This criterion may be achieved on those slowly rotating planets where magma oceans occur due to giant impacts \citep{elkins2012magma}.     

When dominated by strong wind forcing, Regime I will only be relevant when vertical viscosity is larger than 0.4 m$^2$\,s$^{-1}$. Meanwhile, the rotation period should be beyond $\sim$30 Earth days with a viscosity of 0.4 m$^2$\,s$^{-1}$. This requirement for viscosity is relatively strict and may be satisfied by the presence of very strong eddy viscosity \citep{luo2023study,sentchev2023estimation}.


\section{Conclusions}\label{sec:conclude}

On tidally locked lava planets, magma ocean may form on the permanent dayside. Circulation of the magma ocean can be driven by the heterogeneity of stellar radiation and the stress created by the atmosphere flow from the hot dayside to the cold nightside \citep{castan2011atmospheres,nguyen2020modelling, kang2021escaping} as sketched in Figure \ref{fig:schematic}. The strength and depth of the ocean circulation follows different scaling laws depending on whether wind forcing or thermal forcing is dominant and on the dominant balance of the momentum equation (Figure \ref{fig:equation}). 
In this study, we explore the controlling factors of the strength of ocean circulation, the depth of magma ocean, and the ocean heat transport convergence driven by stellar and wind forcings in three dynamic regimes: non-rotating viscosity-dominant Regime I, non-rotating inviscid limit Regime II, and rotation-dominant Regime III. From Regimes I to III, the pressure gradient force is dominantly balanced by vertical viscosity, nonlinear advection, and Coriolis force, respectively. 

When the main driving force of the ocean circulation is thermal forcing, the magma ocean depth, horizontal velocity, and ocean heat transport convergence scale with vertical diffusivity, vertical viscosity, planetary rotation rate, and surface temperature/buoyancy difference. Utilizing the dominant momentum balance, mass continuity, and advection-diffusion balance, we obtain scaling laws for each of the three dynamic regimes, which are summarized in Table \ref{tab:scaling_sumamry}. 
When the main driving force is wind stress, the interior velocity, no longer internal-determined, but induced by the horizontal convergence/divergence of mass transport in the wind-driven boundary layer. Utilizing the dominant momentum balance, mass continuity, and the boundary layer transport, we obtain scaling laws for ocean depth, current speed, and ocean heat transport convergence for the three dynamic regimes. The results are summarized in Table \ref{tab:scaling_sumamry}. For each scaling law, we determine its applicable conditions in Section \ref{sec:parameter_regime}. For Regime I only, we conducted a set of 2D numerical simulations. The numerical results match scaling prediction reasonably well. The numerical examination of Regime II and Regime III is left for future work.

Following these scaling laws, we examined the sensitivity of magma ocean depth, current speed, and ocean heat transport convergence to various controlling parameters, including the planetary rotation rate, substellar temperature, wind stress amplitude and diffusivity/viscosity. Substituting parameters for a typical lava super-Earth, we found the rotation-dominant Regime III to be the most relevant. Scaling laws predict a magma ocean depth that ranges from a few meters to a few hundred meters deep and an ocean heat transport convergence that is smaller than the stellar insolation by 1--4 orders of magnitude, in line with previous work by \citet{kite2016atmosphere} and Part I. 

It should be noted that, although we treat wind stress, thermal forcing, rotation rate, and diffusivity/viscosity as independent parameters in this work, they are intrinsically related. For example, wind stress $\tau$ is proportional to the square of the atmospheric flow speed, which in turn scales with the substellar temperature $\sqrt{T_{sub}}$ \citep{kang2021escaping}. This means that $\tau$ should scale with $T_{sub}$. In turn, hotter planets usually surround its host star at a closer distance, which yields a faster rotation rate, assuming star luminosity is fixed. As the distance between star and planet decreases, the tidal forcing is also likely to increase, which may induce stronger eddy diffusivity and viscosity \citep{bell1975a, smithyoung2001}. 

\section*{Acknowledgments}
We express our gratitude to Feng Ding for his insightful discussions. J.Y. is supported by NSFC under grant nos. 42075046, 42275134 and 42161144011. This work has been supported by the science research grants from the China Manned Space Project (No. CMS-CSST-2021-B09). W.K. is supported by the MIT startup fund. 

The 2D model employed in this study is developed by the authors and is available at \url{https://github.com/YanhongLai/Two-dimensional-model-for-magma-ocean-on-lava-worlds.git}. The simulation data used are archived at \url{https://doi.org/10.5281/zenodo.11467142}.

\bibliography{main}{}
\bibliographystyle{aasjournal}

\end{document}